\documentclass[11pt,a4paper]{article}

\usepackage[backend=bibtex, style=numeric-comp, sorting=none, doi=false, url=false]{biblatex} 
\usepackage{amsfonts}
\usepackage{graphicx}
\usepackage{subfigure}
\usepackage{bm}
\usepackage{xcolor}
\usepackage[normalem]{ulem}
\usepackage{mathtools}
\usepackage{mathrsfs}
\usepackage{soul}
\usepackage{authblk}
\usepackage{epstopdf}
\usepackage[a4paper, left=1.5cm, right=1.5cm, top=2cm, bottom=2cm]{geometry}

\usepackage{etoolbox}

\DeclarePairedDelimiterX\braket[2]{\langle}{\rangle}{#1 \delimsize\vert #2}
\DeclarePairedDelimiterX\inner[2]{\langle}{\rangle}{#1,#2}
\DeclarePairedDelimiterX\ketbra[2]{\lvert}{\rvert}{#1 \delimsize\rangle \delimsize\langle #2}

\newcommand{\me}[1]{\left\langle #1 \right\rangle }

\makeatletter
\providecommand{\@LN@col}[2]{}
\makeatother

\title{Self-consistent harmonic approximation
  with non-local couplings}
\author[1]{Guido Giachetti}
\author[2]{Nicol\`{o} Defenu}
\author[1,3]{Stefano Ruffo}
\author[1,4]{Andrea Trombettoni}
\affil[1]{SISSA and INFN, Sezione di Trieste, Via Bonomea 265, I-34136 
Trieste, Italy}
\affil[2]{Institute for Theoretical Physics, ETH Z\"urich Wolfgang-Pauli-Str. 27, 8093 Zurich, Switzerland}
\affil[3]{Istituto dei Sistemi Complessi, Consiglio Nazionale delle Ricerche,
Via Madonna del Piano 10, I-50019 Sesto Fiorentino, Italy}
\affil[4]{Department of Physics, University of Trieste, Strada Costiera}

\begin{document}
\maketitle

\begin{abstract}
We derive the self-consistent harmonic approximation for the $2D$ XY model with non-local interactions. The resulting equation for the variational couplings holds for any form of the spin-spin coupling as well as for any dimension. Our analysis is then specialized to power-law couplings decaying with the distance $r$ as $\propto 1/r^{2+\sigma}$  in order to investigate the robustness, at finite $\sigma$, of the Berezinskii-Kosterlitz-Thouless (BKT) transition, which occurs in the short-range limit $\sigma \to \infty$. We propose an ansatz for the functional form of the variational couplings and show that for any $\sigma>2$ the BKT mechanism occurs. The present investigation provides an upper bound for the lower critical threshold $\sigma^\ast=2$, above which the traditional BKT transition persists in spite of the LR couplings.
\end{abstract}

\section{Introduction}
Among the different strategies commonly employed to study
interacting systems, one that is often used -- in its simplest form --
is based on the determination of
the non-interacting model that better approximates, in a variational sense, 
the initial, %difficult to directly solve, 
interacting problem. This method, referred to as
the Self-Consistent Harmonic Approximation (SCHA), can be improved
by considering non-quadratic, but solvable,
approximations of the problem at hand
or by an integration of the quantum fluctuations to determine
an optimized classical potential \cite{Koehler1966, Cuccoli_1995}. Then, the SCHA and its variants 
may take different forms both for classical or quantum models, and for equilibrium or dynamical properties. They are in general based on the variational principle of the minimization
of the energy (or free energy) difference between the interacting model
and the approximating one, calculated using the state of the latter \cite{Feynman}. When the approximating model is non-interacting, this approach cannot be used to determine non-trivial  correlation functions or
interference effects, but also in this case SCHA provides estimates for the
equilibrium free energy and the properties of the phases of the model under study.
SCHA and its variants, including the so-called pure quantum SCHA, have been as well used to
provide an estimate of the quantum corrections to the free energy of
nonlinear systems \cite{Giachetti85,Feynman86} and it has proven useful in several contexts, ranging from the study of phonon spectra in
metals and insulators (see \cite{errea} and refs. therein) to
quantum antiferromagnets \cite{Cuccoli_1997} and arrays of Josephson junctions
\cite{Simanek}.

A context in which SCHA has been widely employed along the
years is provided by the two-dimensional XY model, where
the celebrated Berezinskii-Kosterlitz-Thouless (BKT)
transition takes place \cite{Berezinskii1972,Kosterlitz1973,Kosterlitz1974}.
In the BKT transition, the low temperature phase features vortex-antivortex
pairs and power-law correlation functions
(with temperature-dependent exponents), while
in the high temperature phase topological charges can freely propagate leading to exponential decay in the correlation functions
\cite{Minnhagen1981,Kenna1997,Gulacsi1998,Kosterlitz2016}.
The superfluid density
exhibits a universal jump at the critical temperature $T_{BKT}$
\cite{Nelson1977} and, correspondingly, both the critical exponent $\eta$
\cite{Kosterlitz1973,Kosterlitz1974} and the power law exponent for the finite size scaling of the largest eigenvalue of the one-body density matrix \cite{Colcelli20}, jump at $T_{BKT}$
from a finite value, respectively $1/4$ and $7/8$, to zero. 

The Hamiltonian of the short-range, nearest-neighbor XY model reads
\begin{equation}\label{original_SR}
  H = - \frac{J_0}{2} \sum_{\mathbf{i},\mathbf{j}}
  \cos ( \theta_{\mathbf{i}} - \theta_{\mathbf{j}}),
\end{equation}
where $J_0$ is the coupling constant and the sum is on all the pairs of
nearest-neighbor sites of a $2D$ square lattice, on which the variables
$\theta_i$ are defined. The
temperature $T_{BKT}$ of the XY model \eqref{original_SR}
has been the subject of considerable work,
and the value determined by Monte Carlo simulations is given by
$k_B T_{BKT}/J_0=0.893 \pm 0.001$
\cite{GuptaPRL,GuptaPRB,Hasenbusch05,Schultka,Komura12,Hsieh2013},
with $k_B$ the Boltzmann constant. We will put,
as usual, $\beta=1/k_B T$,
with $T$ the temperature.

The SCHA for the short-range Hamiltonian \eqref{original_SR}
applies in a particularly transparent
way. One introduces the quadratic Hamiltonian
$H_0 = \frac{\tilde{J}_0}{4} \sum_{\mathbf{i},\mathbf{j}}
( \theta_{\mathbf{i}} - \theta_{\mathbf{j}})^2$
and then variationally determines
$\tilde{J}_0$ as a function of $J_0$ and $T$, or -- equivalently --
the dimensionless coupling $\beta \tilde{J}_0$ as a function of $\beta J_0$
\cite{PU,Lozovik81,Pires1993,Pires1996}. It is found that $\tilde{J}_0$
is different from zero for $T$ smaller than a temperature, denoted by
$T_c$, at which the effective $\tilde{J}_0$ drops to zero \cite{PU}. An
improved determination of the BKT critical temperature $T_{BKT}$ can be
obtained inserting $\tilde{J}_0(T)$ in the Nelson-Kosterlitz condition \cite{Nelson1977}
at the BKT critical point 
\cite{Lozovik81,Pires1993}. An advantage of this approach is that it can be
extended to the quantum phase model \cite{Simanek,Cuccoli2000,Smerzi2004}
describing arrays of Josephson junctions \cite{Martinoli2000,Fazio2001} and
ultracold bosons in optical lattices in the large filling limit \cite{Trombettoni2005}.

When the couplings are non-local, i.e.
the spins in sites $\mathbf{i}$ and $\mathbf{j}$ are coupled
with a strength $J(\mathbf{i}, \mathbf{j})$, 
the application of the SCHA with the introduction
of a variational coupling matrix $\tilde{J}(\mathbf{i}, \mathbf{j})$
faces with the practical problem of solving the full set of conditions
for the $\tilde{J}$'s. Here, we focus on the case
of Hamiltonians with non-local couplings, proposing and discussing
the consequences of a functional ansatz for the variational couplings
$\tilde{J}(\mathbf{i}, \mathbf{j})$. In particular, we will consider the case of the $d=2$
XY with power-law couplings $J(\mathbf{i}, \mathbf{j})$
decaying asymptotically as $J \sim 1/|\mathbf{i} - \mathbf{j}|^{2+ \sigma}$. The reason for this choice is three-fold.

{\it i)} The
nearest-neighbor $2D$ XY model has been extensively studied with SCHA
\cite{PU,Lozovik81,Pires1993,Pires1996} providing a benchmark for the
short-range limit of the theory. {\it ii)}
The study of statistical mechanics
models with long-range interactions attracted considerable attention along the
last few decades (see the reviews \cite{Campa2009,campa2014physics,Defenu_2020}).
The general result is that exists a value of the exponent of the power-law decay, denoted
by $\sigma^\ast$, such that for $\sigma>\sigma^\ast$ the universality
class is the same of the short-range limit $\sigma \to \infty$. A
first derivation by Sak, focusing on $O(n)$ models \cite{sak1973recursion}, provided the result $\sigma^\ast=2-\eta_{SR}$, where $\eta_{SR}$ is the anomalous
dimension of the short-range ($\sigma\to\infty$) limit. The validity of this result has been
thoroughly investigated using a variety of techniques
\cite{Luijten1997,Luijten2002,blanchard2013influence,Brezin:2014bl,Angelini2014,Defenu2015,Defenu2016b,Behan2017,Behan2017b,Gori2017,Horita2016}, yielding very strong evidences in its favor. Nevertheless, Sak's result only concerns second-order phase transitions, while the BKT transition is an infinite order one, so that the traditional $\sigma^{*}=2-\eta_{SR}$ threshold does not apply to the XY model with non-local couplings.
{\it iii)} Results on the BKT transition for the $2D$ XY model
with non-local couplings are very rare, to the best of
our knowledge, with exceptions coming from related models such as
$1D$ quantum XXZ spin models
with long-range interactions
\cite{Bermudez2016,Maghrebi2017,botzung2019effects}
and random-dilute graphs \cite{Cescatti2019}.
Then, qualitative information coming
from the SCHA may provide insights for further investigations.
\section{SCHA with non-local couplings} We consider the XY model on a $2$-dimensional square lattice of $N$ sites with non-local couplings
\begin{equation}\label{original}
\beta H \equiv - \frac{1}{2} \sum_{\mathbf{i},\mathbf{j}} J(r) \cos ( \theta_{\mathbf{i}} - \theta_{\mathbf{j}}).
\end{equation}
where $r = |\mathbf{r}| $ and $\mathbf{r} = \mathbf{i}-\mathbf{j}$ and, as previously mentioned, we choose
\begin{equation}
  J(r) = \frac{J}{r^{2+ \sigma}},
  \label{J:LR}
\end{equation}
with $\sigma>0$.
In Eq.\,\eqref{original} we are setting $J \equiv \beta J_0$ and
we take energy in units of $J_0$ (restoring $J_0$
when useful for clarity).

However, for the moment we do not need to fix a specific form for the couplings
$J(r)$, and our results till Eq.\,\eqref{Min} hold for any non-local couplings
$J(r)$.

%Thus is natural to think to the SCHA as a useful tool to gain some information about the system.
Proceeding according to the SCHA, we 
replace the cosine in the original Hamiltonian \eqref{original} with a quadratic term
\begin{equation} \label{ansatz}
\beta H_0 =  \frac{1}{4} \sum_{\mathbf{i},\mathbf{j}} \tilde{J}(\mathbf{r}) ( \theta_{\mathbf{i}} - \theta_{\mathbf{j}})^2
\end{equation}
where $\tilde{J}(\mathbf{r})$ is a generic function that has to be determined in a self-consistent
way, in order to approximate the original Hamiltonian at best. One may wonder whether the the asymptotic behavior of $\tilde{J}(\mathbf{i})$ can be
different from that of $J(r)$, thus it is convenient to derive the self-consistent equations in the most general setting.
%Assuming then $\tilde{J}(r) \sim r^{-2-\tilde{\sigma}}$ with $\tilde{\sigma} \neq \sigma$ in general we notice that the phenomenology described by \eqref{ansatz} can be non trivial. Indeed the continuous limit of Eq. \eqref{ansatz} depends on the value of $\tilde{\sigma}$; in particular for $\tilde{\sigma}>2$ we recover a short range coupling. Had we fixed $\tilde{\sigma} = \sigma$ in our ansatz, then we could expect the border the long range behavior to take over for $\sigma<2$. In this case, however, even for $\sigma>2$ new physics could reasonably  arise. 

%\subsection{Variational free energy}
To further proceed, we introduce the partition function of the quadratic model 
\begin{equation}
Z_0 = \int \prod_{\mathbf{j}} d \theta_{\mathbf{j}} e^{- \beta H_0},
\end{equation}
and the corresponding free energy $F_0$ defined by $Z_0= e^{- \beta F_0}$.
The average on the corresponding Boltzmann measure is defined as
\begin{equation}
\me{\cdot}_0 = \frac{1}{Z_0} \int \prod_{\mathbf{j}} d \theta_{\mathbf{j}} e^{- \beta H_0}.
\end{equation}
In agreement with the variational principle \cite{Feynman}
the best possible result for the couplings $\tilde{J}$ is obtained by minimizing the quantity
\begin{equation}
\mathcal{F} = \beta F_0 + \beta \me{H}_0 - \beta \me{H_0}_0
\end{equation}
with respect to $\tilde{J} (\mathbf{r})$.
The equipartition theorem implies
that $\me{H_0}_0 = \frac{N}{2 \beta}$, so that this term can be ignored. On the other hand: 
\begin{equation}
\begin{split}
  \beta \me{H}_0 &= - \frac{1}{2} \sum_{\mathbf{i},\mathbf{j}} J(|\mathbf{i}-\mathbf{j}|) \me{\cos ( \theta_{\mathbf{j}} - \theta_{\mathbf{j}})}_0 \\
  %&= - \frac{1}{2} \sum_{\mathbf{i},\mathbf{j}} J(|\mathbf{i}-\mathbf{j}|) e^{- \frac{1}{2}\me{( \theta_{\mathbf{j}} - \theta_{\mathbf{j}})^2}_0} \\
  &= - \frac{N}{2} \sum_{\mathbf{r}} J(r) e^{- \frac{1}{2}\me{( \theta_{0} - \theta_{\mathbf{r}})^2}_0} 
\end{split}
\end{equation}
where we made use of the identity $\me{\cos A}_0 = e^{- \frac{1}{2} \me{A^2}_0}$,
valid for every Gaussian measure, and of the translational invariance of the system. Finally, in
order to find $F_0$, and to compute the correlation functions, we have to diagonalize $H_0$ by means
of the Fourier transform. Defining
\begin{equation}
  \theta_{\mathbf{q}} = \frac{1}{\sqrt{N}} \sum_{\mathbf{j}} e^{ - i \mathbf{q} \cdot \mathbf{j}} \ \theta_{\mathbf{j}} \, ; \hspace{0.5cm} \theta_{\mathbf{j}} = \frac{1}{\sqrt{N}} \sum_{\mathbf{q} \in BZ} e^{  i \mathbf{q} \cdot \mathbf{j}} \ \theta_{\mathbf{q}}
\end{equation}
(where the sum on the wavevectors $\mathbf{q}$ is on the first Brillouin zone) we then have 
\begin{equation} \label{diagH0}
\beta H_0 = \frac{1}{2} \sum_{\mathbf{q} \in BZ} K(\mathbf{q}) |\theta_{\mathbf{q}}|^2
\end{equation}
where
\begin{equation} \label{Kq}
K(\mathbf{q}) = \sum_{\mathbf{r}} \tilde{J} (\mathbf{r}) \Big( 1 - \cos(\mathbf{q} \cdot \mathbf{r}) \Big) 
\end{equation}
%Then it is easy to see that
It follows $Z_0 = (2 \pi)^{N/2} \prod_{\mathbf{q} \in BZ} K(\mathbf{q})^{-1/2}$ and
\begin{equation}
\beta F_0 = \frac{1}{2} \sum_{\mathbf{q} \in BZ} \ln K(\mathbf{q})
\end{equation}
From Eq.\,\eqref{diagH0} follows that $\me{\theta_\mathbf{q}\theta_\mathbf{q'}}_0 = \delta_{\mathbf{q} + \mathbf{q'}} K(\mathbf{q})^{-1}$. Then
\begin{equation}
\begin{split}
\me{( \theta_{\mathbf{0}} - \theta_{\mathbf{r}})^2}_0 &= \frac{1}{N} \sum_{\mathbf{q},\mathbf{q'} \in BZ} \me{\theta_\mathbf{q}\theta_\mathbf{q'}}_0 ( 1 - e^{i \mathbf{q} \cdot \mathbf{r}})( 1 - e^{i \mathbf{q'} \cdot \mathbf{r}}) \\ &= \frac{2}{N} \sum_{\mathbf{q} \in BZ} \frac{1 - \cos(\mathbf{q} \cdot \mathbf{r})}{K(\mathbf{q})}
\end{split}
\end{equation}

\section{%Minimization of $\mathcal{F}$
  %Equation for the couplings $\tilde{J}$
Minimization of the free energy}
We finally find
\begin{equation} \label{Fvar}
  \mathcal{F} = \frac{1}{2} \sum_{\mathbf{q} \in BZ} \ln K(\mathbf{q}) -
  \frac{N}{2} \sum_{\mathbf{r}} J(r) e^{- G(\mathbf{r})}
\end{equation}
with
\begin{equation} \label{Gr}
  G(\mathbf{r}) = \frac{1}{N} \sum_{\mathbf{q} \in BZ} \frac{1 - \cos(\mathbf{q} \cdot \mathbf{r})}{K(\mathbf{q})}.
\end{equation}
In the lattice case and at generic distance $\mathbf{r}$, the quantity
$G(\mathbf{r})$ also depends on the direction of the vector ${\mathbf r}$, but for large distances
one may see that it only depends on the modulus $r$, as expected.

A close inspection of Eq.\,\eqref{Fvar} reveals that the couplings $\tilde{J} (\mathbf{r})$ appear in
$\mathcal{F}$ only through the quantities $K(\mathbf{q})$.
Then, to proceed with the minimization of $\mathcal{F}$,
it is sufficient to derive with respect to the $K(\mathbf{q})$'s, obtaining
\begin{equation}
  %\pd{\mathcal{F}}{K(\mathbf{q})} = \frac{1}{2 K(\mathbf{q})} + \frac{N}{2} \sum_{\mathbf{r}} J(r) \pd{G(r)}{K(q)} e^{-G(r)}
  \frac{\delta \mathcal{F}}{\delta K(\mathbf{q})} =
  \frac{1}{2 K(\mathbf{q})} + \frac{N}{2} \sum_{\mathbf{r}} J(r)
  \frac{\delta G(\mathbf{r})}{\delta K(\mathbf{q})} e^{-G(\mathbf{r})}  
\end{equation}
and in turn
\begin{equation}
  %\pd{G(r)}{K(q)} = - \frac{1}{N} \frac{1 - \cos( \mathbf{q} \cdot \mathbf{r})}{K(q)^2}
  \frac{\delta G(\mathbf{r})}{\delta K(\mathbf{q})} = - \frac{1}{N} \frac{1 - \cos( \mathbf{q} \cdot \mathbf{r})}{K(\mathbf{q})^2}, 
\end{equation}
so that
\begin{equation}
  %\pd{\mathcal{F}}{K(\mathbf{q})} = \frac{1}{2 K(q)^2} \left( K(q) - \sum_r J(r) \left( 1 - \cos (\mathbf{q} \cdot \mathbf{r}) \right) e^{-G(r)} \right)
  \frac{\delta \mathcal{F}}{\delta K(\mathbf{q})} =
  \frac{K(\mathbf{q}) - \sum_{\mathbf{r}} J(r) \left[ 1 - \cos (\mathbf{q} \cdot \mathbf{r}) \right] e^{-G(\mathbf{r})}}{2 K(\mathbf{q})^2}. 
\end{equation}
Exploiting the definition \eqref{Kq} for $K(\mathbf{q})$ it follows 
\begin{equation}
  %\pd{\mathcal{F}}{K(\mathbf{q})} = \frac{1}{2 K(q)^2}  \sum_{\mathbf{r}} \left( \tilde{J}(\mathbf{r}) - J(r) e^{-G(\mathbf{r})} \right) \left( 1 - \cos (\mathbf{q} \cdot \mathbf{r}) \right)
  \frac{\delta \mathcal{F}}{\delta K(\mathbf{q})} = \frac{1}{2 K(\mathbf{q})^2}  \sum_{\mathbf{r}} \mathcal{A}(\mathbf{r}) \Big( 1 - \cos (\mathbf{q} \cdot \mathbf{r}) \Big),
  \label{CalA}
\end{equation}
where $\mathcal{A}(\mathbf{r}) \equiv \tilde{J}(\mathbf{r}) - J(r) e^{-G(\mathbf{r})}$,
Eq.\,\eqref{CalA} being valid for each value of $\mathbf{q} \in BZ$ and implying: 
\begin{equation} \label{Min}
J(r) = \tilde{J}(\mathbf{r}) e^{G(\mathbf{r})}.
\end{equation}

Eq.\,\eqref{Min} is the desired relation between the couplings $\tilde{J}(\mathbf{r})$ of the optimizing model and the couplings $J(r)$ of the initial model.
In solving it, one can actually look for a solution such
that the couplings $\tilde{J}$ depend only on $r$. Eq.\,\eqref{Min} has been
derived for the $2D$ XY model, but the same calculations can be extended for different dimensions. The same structure of Eq.\,\eqref{Min} is found for $O(n)$ models as well. Finally, we notice that Eq.\,\eqref{Min} is valid
for any non-local form of the couplings $J(r)$, such as the exponential one, and, therefore, it is not limited the power-law decaying form in Eq.\,\eqref{J:LR}.

\begin{figure} 
    \centering
    \includegraphics[scale=0.64]{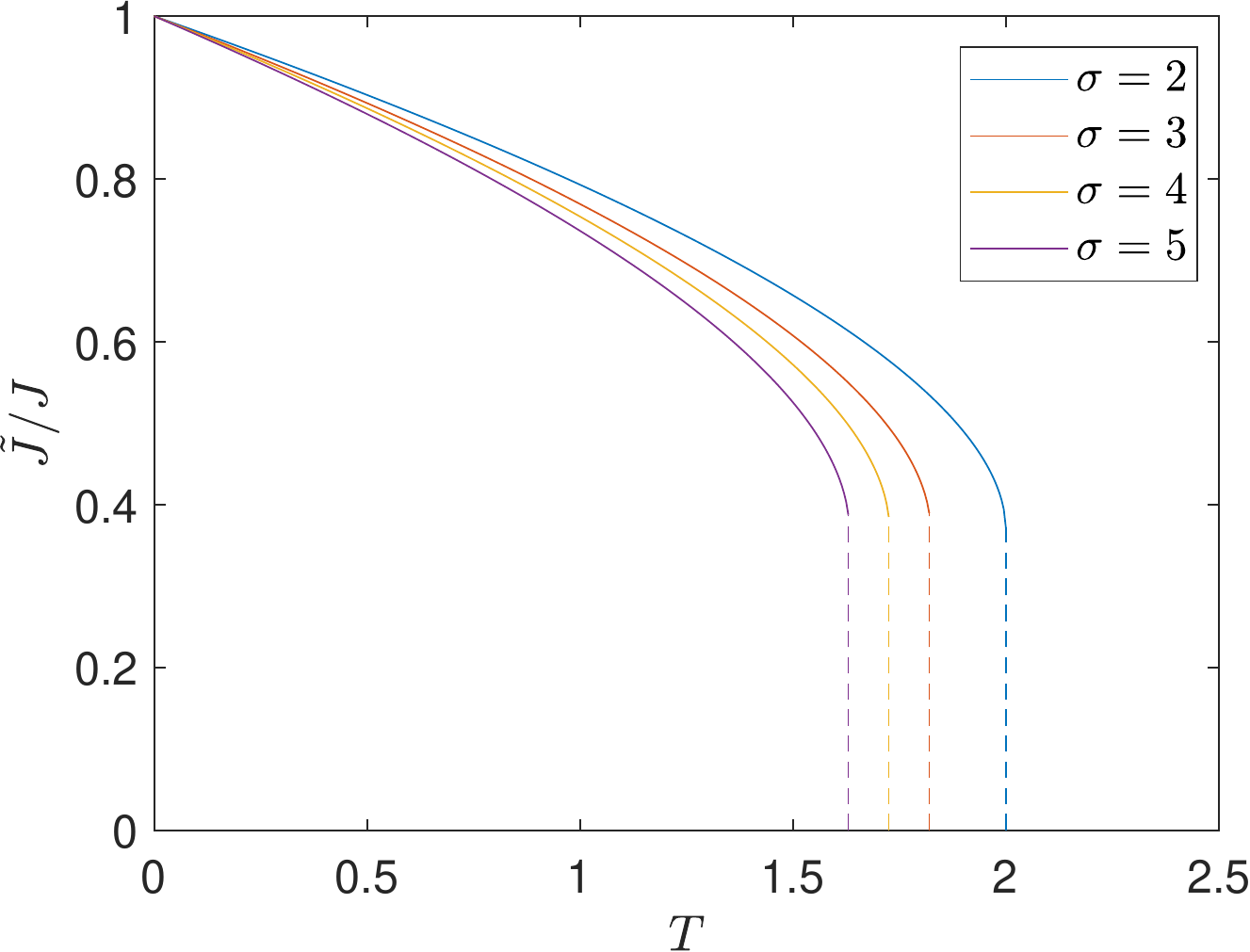}
    \caption{Ratio $\tilde{J}/J$ as a function of $T = 1/J$
      for different values of $\sigma$. For each $\sigma$ we find a jump
      of $\tilde{J}$ to zero at a temperature denoted in the text
      and in Fig. \ref{fig:Tcs} as $T_c$.}
    \label{fig:JtildeJ}
\end{figure}

\begin{figure} 
    \centering
    \includegraphics[scale=0.64]{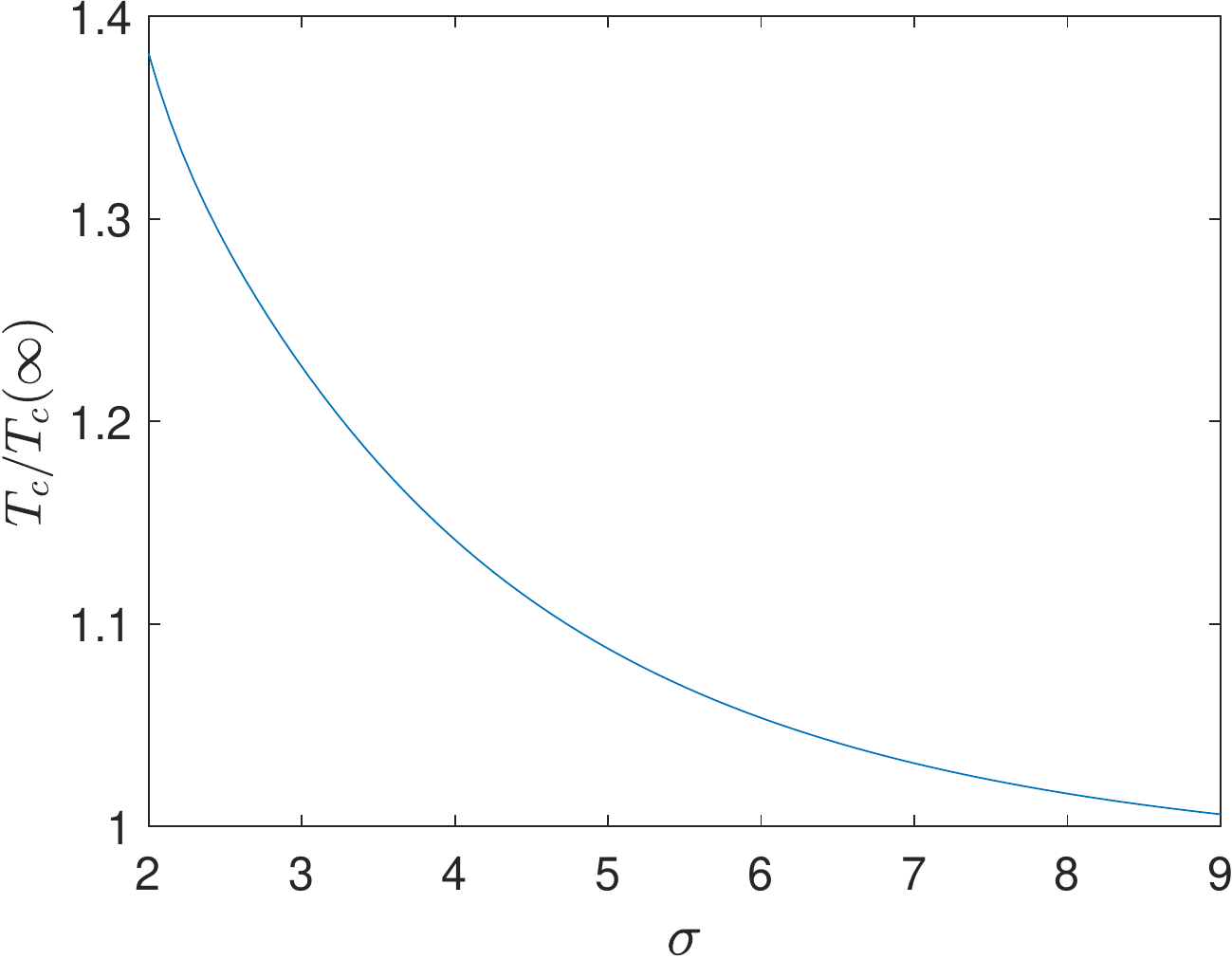}
    \caption{$T_{c} (\sigma)$ at which $\tilde{J}$ jumps, in units
      of the $\sigma \rightarrow \infty$ value
      $T_{c} (\infty) = \frac{4}{e}$. We see that the temperature is
      finite for $\tilde{\sigma} \rightarrow 2$.}
    \label{fig:Tcs}
\end{figure}

\begin{figure} 
    \centering
    \includegraphics[scale=0.64]{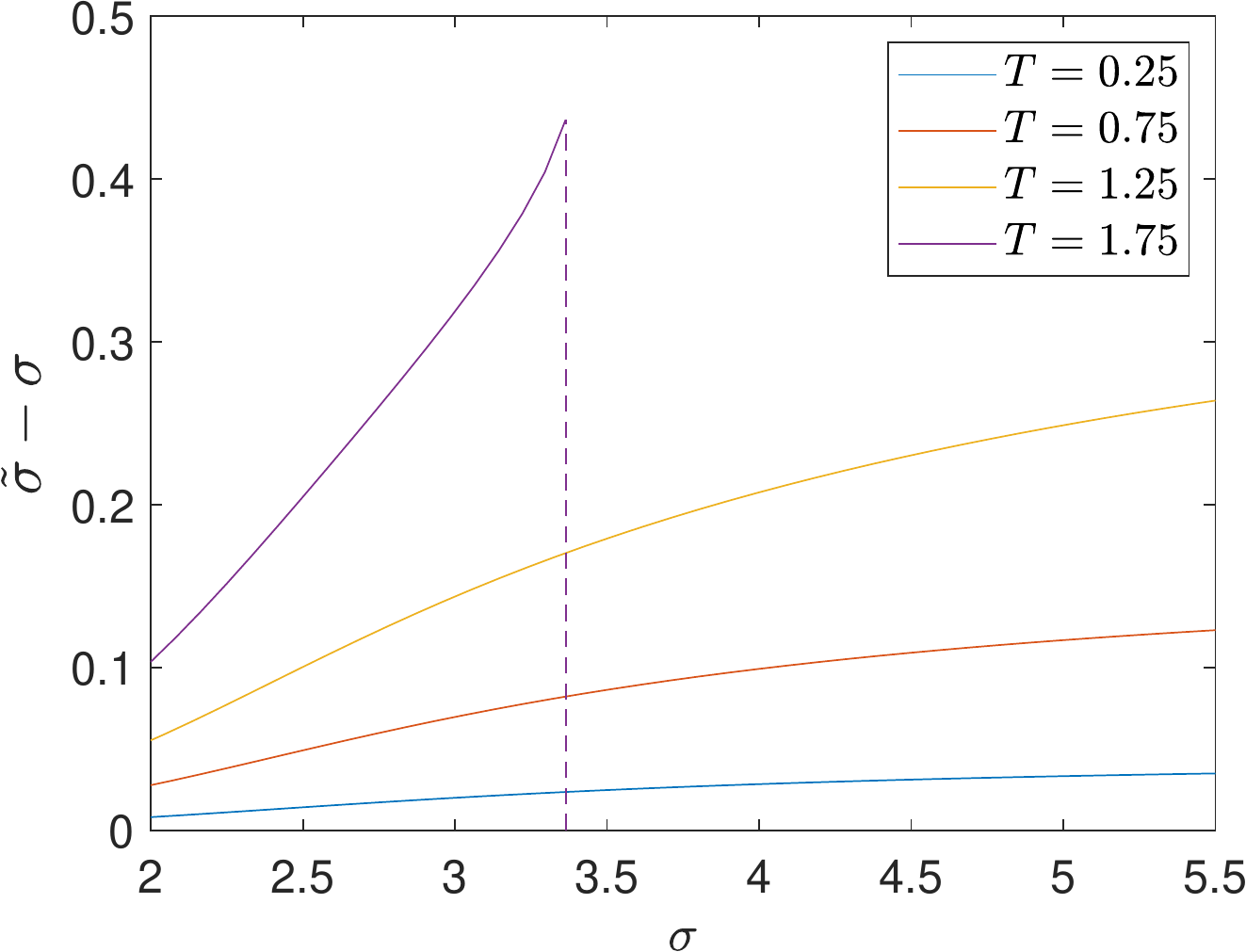}
    \caption{Behavior of $\tilde{\sigma}-\sigma$ as a
      function of $\sigma$ for different values of the temperature $T$.
      It is seen that $\tilde{\sigma} > \sigma$ everywhere.}
    \label{fig:deltasigma}
\end{figure}

\section{Ansatz for the variational couplings}
We now come back to the specific problem of the $2D$ XY model with
power-law decaying couplings, as defined in Eq.\,\eqref{J:LR}: $J(r) = \frac{J}{r^{2+ \sigma}}$.
We remind that we are setting $J = \beta J_0$ and
energy units so that $J_0=1$ (unless differently stated).

Within this choice, when the parameter
$\sigma \to \infty$, only nearest-neighbor couplings are present, and the 
Hamiltonian \eqref{original_SR} with coupling $J_0$ between pairs of
nearest-neighbors is retrieved.
Moreover, we assume $\sigma >0$
so that the additivity of the thermodynamic quantities
is preserved \cite{Campa2009,campa2014physics}.
As recalled in the Introduction, in the short-range limit
$\sigma \rightarrow \infty$, the system undergoes the BKT transition and this phenomenology is qualitatively
captured by the SCHA approximation, in which
the original Hamiltonian is replaced with a
variational quadratic ansatz. The variational
coupling jumps discontinuously to zero at a temperature $T_c$ (see Fig.\,\ref{fig:JtildeJ}) and
provides an estimate for the superfluid stiffness, which can be
used in the BKT renormalization group flow equations
\cite{PU,Lozovik81,Pires1993,Pires1996}.
%and this implies that the exponent of the correlations goes from a finite value to infinity. 

Our main goal is to discuss the corresponding behavior for finite
$\sigma$. Without being able to
clarify the nature of the universality class and the value of critical
exponents, SCHA is
nevertheless giving first information whether the BKT phenomenology is stable
for large $\sigma$, as one
would expect, and if one can perform an estimate, or -- better --
put an upper bound for the value of
$\sigma^{\ast}$ (defined in this $d=2$ $O(2)$ case such that for $\sigma>\sigma^\ast$ one has a BKT transition).

For large $\sigma$ (i.e. in the short-range limit),
the quantity $\me{\cos (\theta_{\mathbf{j+r}}- \theta_{\mathbf{j}})}_0 =  e^{-G(\mathbf{r})}$ has a temperature-dependent power law behavior. In order to study such temperature dependence in the power-law case, one needs to extract
the large distance properties of the variational couplings
$\tilde{J}(\mathbf{r})$. Therefore, it is natural  to study
how a different asymptotic power-law behavior in $\tilde{J}(\mathbf{r})$ with
respect to $J(r)$ can actually arise from \eqref{Min}, as anticipated.
This lead to the ansatz 
\begin{equation}
\tilde{J} (\mathbf{r}) \equiv \frac{\tilde{J}}{r^{2 + \tilde{\sigma}}}
\label{Ansa}
\end{equation}
where $\tilde{\sigma}$ may, in general, be different from $\sigma$. Inserting the ansatz
\eqref{Ansa} in \eqref{Min}, we have to determine
$\tilde{\sigma}$ and $\tilde{J}$ as a function of $\sigma$ and $J$. We know that in the short-range limit
($\tilde{\sigma}, \sigma \rightarrow \infty$)
$\tilde{J}$ as a function of $T$ has a jump from a finite value to zero
\cite{PU}.

An advantage of the ansatz \eqref{Ansa} is to give
first information about the robustness of the BKT transition
from the knowledge of $\tilde{\sigma}$. Indeed, in the short-range
nearest-neighbor case, a textbook calculation gives an estimate of $T_{BKT}$
by calculating the energetic and entropic contributions to the free energy
$\Delta F$ of a free vortex, see e.g. Chapter 4 of \cite{LeBellac}. By putting
$N=L^2$ and the lattice spacing $\equiv 1$, one has that $\Delta F=\pi J_0
\ln{L}-k_B T \ln{L^2}$, giving $k_B T_{BKT}=\pi J_0/2$. A similar situation
of competition between the energetic and entropic contributions 
occurs for the $1D$ Ising model with power-law interactions
($\propto |r|^{-1-\sigma}$), where the excitations are magnetization kinks rather than vortices
\cite{Thouless69}. In our $2D$ case, one can see that if
$\tilde{\sigma}>2$ then BKT behavior is expected. This can be understood
by observing that for $q \to 0$ the scaling of the
propagator due to the effective interactions ($\propto 1/q^{\tilde{\sigma}}$) is irrelevant with respect to the one of the free propagator ($\propto 1/q^2$), from which one can conclude
the irrelevance of the non-local interaction when $\tilde{\sigma}>2$. We notice
that this is in agreement with the known rigorous result of a
low-temperature phase with spontaneous symmetry breaking
and magnetization for $XY$ couplings decaying faster than
$1/r^4$ in $2D$ \cite{kunz1976}. In conclusion, if one finds
\begin{equation}
\tilde{\sigma}(\sigma)>2,
\label{cond_BKT}
\end{equation}
then it is possible to conclude that SCHA is indicating persistence of
BKT at that $\sigma$.

\begin{figure} 
    \centering
    \includegraphics[scale=0.64]{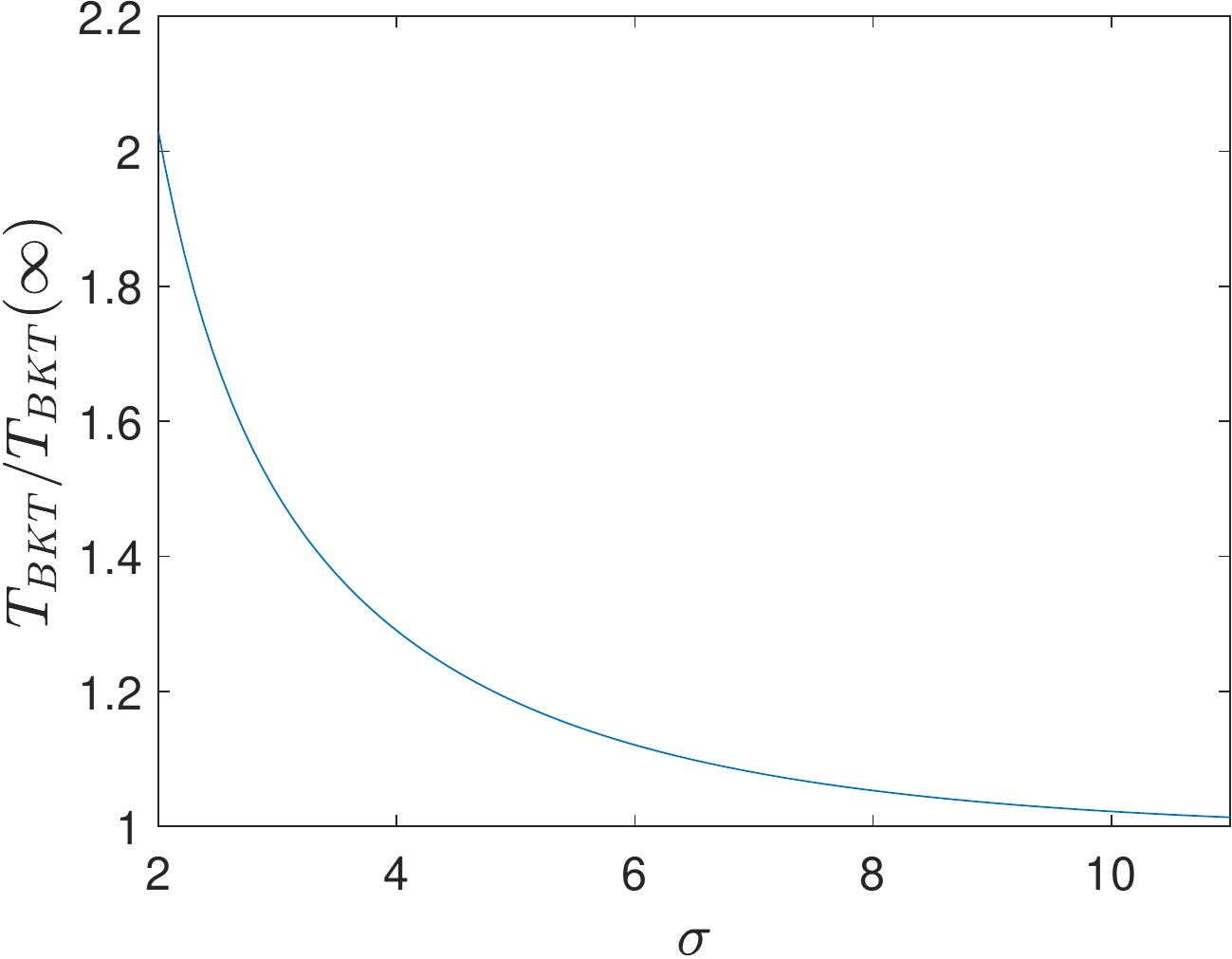}
    \caption{Estimate of $T_{BKT} (\sigma)$
      in units of the $\sigma \rightarrow \infty$ value
      $T_{BKT} (\infty)$.
      %We see the temperature is finite as $\tilde{\sigma} \rightarrow 2$.
    }
    \label{fig:TBKT}
\end{figure}

\section{Results}
Let us start from the case $\sigma>2$, deferring our comments on the case
$\sigma<2$ to the end of the present Section.
By plugging the ansatz \eqref{Ansa} in Eq.\,\eqref{Min}, we numerically
solve for $\tilde{J}$ and $\tilde{\sigma}$ in a square lattice. We considered
growing values of $N$ to solve Eq.\,\eqref{Min}, and we observed that
convergence of the result is obtained for $N \sim 10^4$.

In Fig. \ref{fig:JtildeJ} we show
$\tilde{J}/J$ as a function of the temperature for different values of
$\sigma \geq 2$. The temperature $T_c$ corresponding to the jump
is $\sigma$-dependent (see below for a discussion of the difference
between $T_c$ and $T_{BKT}$). In the $\sigma \rightarrow \infty$ limit
it converges to the short-range value $k_B T_c(\infty) = \frac{4}{e} J_0$
\cite{PU}. The behavior of $T_c(\sigma)$ in units of $T_c(\infty)$
is shown in Fig. \ref{fig:Tcs}. We found that $T_{c}$ is a
decreasing function of $\sigma$ which remains finite
as $\sigma$ approaches $2$.

Our main result is that one has $\tilde{\sigma}>2$.
This is apparent from Fig \ref{fig:deltasigma}, where we show
the behavior of $\tilde{\sigma} - \sigma$ as a function of $\sigma$
for different values of the temperature $T$.
This quantity is of course only defined as long as $T < T_c(\sigma)$.
In particular, since $T_c(\sigma) \rightarrow T_c(\infty)$ as
$\sigma \rightarrow \infty$, for $T<T_{c} (\infty)$ the curve is defined
for every $\sigma$. As expected, $\tilde{\sigma} - \sigma$
is temperature-dependent and $\tilde{\sigma} \rightarrow \sigma$ as
$T \rightarrow 0$. For $T<T_c (\infty)$, as
$\sigma \rightarrow \infty$, $\tilde{\sigma} - \sigma$ goes to a constant, in agreement with the natural expectation of recovering the short-range result.
It is important to notice that, for each value of the parameters,
we have that $\tilde{\sigma} > \sigma$ and therefore the validity of the
condition \eqref{cond_BKT}.

Going to the $\sigma<2$ case, since
$T_c$ is finite for $\sigma \rightarrow 2^{+}$ and also
$\tilde{\sigma} > \sigma$, one may be tempted to investigate whether
the BKT phase can survive in the $\sigma<2$ regime, and
ascertain if the value of $\sigma$ for which the long-range tail
of the couplings modifies the critical behavior is actually smaller
than $2$. However, for $\sigma<2$ the solution of Eq.\,\eqref{Min}
with $\tilde{J}(\mathbf{r})$ in the form of Eq.\,\eqref{Ansa} is no longer unique. Then, 
since our ansatz considers only a subspace of the possible functional
forms of $\tilde{J}(r)$ and the difference in the energy of the solutions
can be of the same order of the error introduced by the ansatz \eqref{Ansa},
we cannot safely draw any conclusion for $\sigma<2$. To further explore
the $\sigma<2$ region one has to resort
both to the numerical solution of Eq.\,\eqref{Min} without the use of  the ansatz \eqref{Ansa} and to the
application of more sophisticated techniques. It is anyway fair
to conclude that $2$ is an upper bound
for $\sigma^\ast$, i.e. for $\sigma>2$ one has a BKT transition at finite
temperature.

\section{Estimate of $T_{BKT}$}
$T_c(\sigma)$ can be considered only
a very crudest estimate of the temperature at which the universal jump occurs.
For the short-range limit, a better estimate of $T_{BKT}$ can be obtained
if we define $T_{BKT}$ as the temperature at which the variational
$\tilde{J}$ reaches the value $\tilde{J}_{BKT} = \frac{2}{\pi}$
\cite{Pires1993}. This value, restoring the temperature, is obtained
by considering the Nelson-Kosterlitz condition \cite{Nelson1977}
for the short-range $XY$ model, reading
$J_{superfl}/k_B T=2/\pi$, and using the SCHA value $\tilde{J}_0$ as estimate
for the superfluid stiffness $J_{superfl}$. In this way, one can obtain
$k_B T_{BKT}=1.06 J_0$, improving with respect to the value
$T_c(\sigma\to\infty)$ (given by $k_B T_{c}=(4/e) J_0=1.47J_0$). A further
improvement could be obtained by using estimates of the dielectric constant in the Nelson-Kosterlitz transition, giving $k_B T_{BKT}=0.96 J_0$
\cite{Pires1993}. These estimates can be compared with the functional
renormalization group estimate, $k_B T_{BKT}=0.94 J_0$ \cite{defenu2017-1}, with a (functional) renormalization group calculation using a renormalized initial condition, $k_B T_{BKT}= 0.89 J_0$\cite{Maccari_2020}, with the analytic calculation based on the mapping on the
$1D$ quantum XXZ model, $k_B T_{BKT}=0.883 J_0$ \cite{Mattis84}, and with
the previously mentioned
Monte Carlo value $k_B T_{BKT}=0.893 J_0$. We notice that the
Nelson-Kosterlitz condition can be used to extract $T_{BKT}$ with high precision
from Monte Carlo data \cite{Hsieh2013}.

The condition $\tilde{J}_{BKT} = \frac{2}{\pi}$ can be thought as
corresponding to the value at which the Gaussian
low-temperature theory becomes unstable due to the presence of the
topological excitations \cite{Kosterlitz1973,Kosterlitz1974}.
This improves the estimate of $T_{BKT}$, since the
plain SCHA does not account for the presence of these excitations,
i.e. the real mechanism underlying the transition.

Then, we expect that a better estimate of $T_{BKT} (\sigma)$ can be obtained
if we apply the same line of reasoning for finite $\sigma$, i.e. using a
two-step approach in which SCHA is used to estimate the superfluid stiffness
and then the latter is used in the Nelson-Kosterlitz condition.
However, with general non-local couplings, and power-law couplings in
particular, a microscopic relation between the coupling $J(r)$ and the
superfluid stiffness is not known. One can think to use $\tilde{J}$ instead
of $J$, but this would heavily underestimate the effect of the tails
of the variational couplings $\tilde{J}$. 

We propose to proceed defining the analog of the nearest-neighbors
coupling in the context of our variational quadratic Hamiltonian
\eqref{Ansa}.
We then make use once again of the variational method to
determine the best value of $J_{nn}$ in
\begin{equation}
  \beta H_{nn} = \frac{J_{nn}}{2} \sum_{\mathbf{i},\mathbf{j}}
  \left( \theta_{\mathbf{i}} - \theta_{\mathbf{j}} \right)^2 
\end{equation}
which best approximates $H_0$. We find the result: 
\begin{equation}
  J_{nn} (\tilde{J}, \tilde{\sigma}) = \frac{1}{2 N}
  \sum_{\mathbf{q} \in BZ} \frac{K(\mathbf{q})}{2-\cos q_x -\cos q_y}
\end{equation}
where $K(\mathbf{q})$ is computed from \eqref{Kq} 
with $\tilde{J}(r) = \frac{\tilde{J}}{r^{2+\sigma}}$ and
$\tilde{J}$, $\tilde{\sigma}$ corresponding to the solutions
of the variational equation \eqref{Min}.

We can then find our estimate of $T_{BKT}$ by choosing
$\tilde{J} = \frac{1}{T}$ such that $J_{nn} = \frac{2}{\pi}$.
The behavior of $T_{BKT}(\sigma)$ in units of the short-range value
$T_{BKT}(\infty)$ are shown in Fig. \ref{fig:TBKT}.
The dependence on $\sigma$ is stronger than those of $T_c(\sigma)$,
even if the qualitative behavior is the same. In particular,
$T_{BKT} (\sigma)$ remains finite as well as
$\sigma \rightarrow 2^{+}$. Whether this suggests or not 
the scenario in which the BKT transition survives for $\sigma < 2$ remains open.

\section{Conclusions} In this paper we studied
the self-consistent harmonic approximation (SCHA)
for the $2D$ XY model with non-local couplings. The approach relies
on the well known procedure of approximating the interacting model under study with a quadratic model,
whose coefficients are optimized minimizing the free energy difference. 
We derived an equation for the variational
couplings which is valid for general non-local interactions
and in a form that can be used for other models, such
as long-range $O(n)$ models. Then, we focused on
power-law couplings $J(r)$ decaying as $J(r)=J/r^{2+\sigma}$.

The short-range limit $\sigma \to \infty$ exhibits
the Berezinskii-Kosterlitz-Thouless (BKT) transition. To extract information
about the robustness of the BKT transition at finite $\sigma$ we propose
an ansatz for the functional form of the variational couplings
$\tilde{J}(r)$ reading $\tilde{J}(r)=\tilde{J}/r^{2+\tilde{\sigma}}$,
where $\tilde{J}$ and $\tilde{\sigma}$
have to be determined as a function of $J$ and $\sigma$.

The study of this dependence revealed that
for $\sigma>2$ BKT occurs. The critical temperature
at which $\tilde{J}$ drops to zero has been determined. We
then used the output of the SCHA calculation in the Nelson-Kosterlitz
condition at the critical point to obtain an improved estimate of the
BKT critical temperature.

Our results suggest that, once the value $\sigma^\ast$ is defined in such a way that for $\sigma>\sigma^\ast$ one has a BKT transition, then
one has an upper bound for $\sigma^\ast$ given by $\sigma^\ast=2$. For $\sigma<2$ the SCHA has two solutions for the variational parameters. Therefore further work is needed
to clarify the structure of the phase diagram and the critical points
for $\sigma<2$.\\

{\it Acknowledgments:} We thank N. Dupuis, T. Enss, G. Gori and I. Nandori for
fruitful discussions. We also thank R. Vaia for useful correspondence. This work is 
supported by the CNR / HAS (Italy-Hungary) project
``Strongly interacting systems in confined geometries'' and  by the Deutsche Forschungsgemeinschaft (DFG, German Research Foundation)  
under Germany’s Excellence Strategy ``EXC-2181/1-
390900948'' (the Heidelberg STRUCTURES Excellence
Cluster). This work is part of MUR-PRIN2017 project ``Coarse-grained description for non- equilibrium systems and transport phenomena (CO-NEST)'' No. 201798CZL whose partial financial support is acknowledged.

\printbibliography[heading=none]

\end{document}